\documentclass[aps,prd,twocolumn,showpacs,groupedaddress,10pt,nofootinbib]{revtex4}
\usepackage{graphicx}
\usepackage{amssymb}
\usepackage{amsmath}
\usepackage{amsfonts}
\usepackage[latin1]{inputenc}
\newcommand{\be}{\begin{equation}}
\newcommand{\ee}{\end{equation}}
\newcommand{\bea}{\begin{eqnarray}}
\newcommand{\eea}{\end{eqnarray}}

\begin{document}

\title{Phase space analysis of quintessence fields trapped in a Randall-Sundrum Braneworld: anisotropic Bianchi I brane with a Positive Dark Radiation term}

\author{Dagoberto Escobar}\email{dagoberto.escobar@reduc.edu.cu}
\affiliation{Departamento de F\'isica Universidad de Camagüey, Cuba.}

\author{Carlos R. Fadragas}\email{fadragas@uclv.edu.cu}
\affiliation{Departamento de F\'isica, Universidad Central de Las
Villas, Santa Clara \enskip CP 54830, Cuba.}

\author{Genly Leon}\email{genly@uclv.edu.cu}
\affiliation{Departamento de Matem\'atica, Universidad Central de Las
Villas, Santa Clara \enskip CP 54830, Cuba.}
\affiliation{Instituto de F\'{\i}sica, Pontificia Universidad de Cat\'olica de Valpara\'{\i}so, Casilla 4950, Valpara\'{\i}so, Chile.}

\author{Yoelsy Leyva}\email{yoelsy.leyva@fisica.ugto.mx}
\affiliation{Divisi\'on de Ciencias e Ingenier\'ia de la Universidad de Guanajuato, A.P. 150, 37150, Le\'on, Guanajuato, M\'exico.}

\date{\today}

\begin{abstract}

In this paper we investigate, from the dynamical systems
perspective, the evolution of an scalar field with arbitrary
potential trapped in a Randall-Sundrum's Braneworld of type 2. We
consider an homogeneous but anisotropic Bianchi I (BI) brane
filled also with a perfect fluid. We also consider the effect of the
projection of the five-dimensional Weyl tensor onto the
three-brane in the form of a positive Dark Radiation term. Using the center manifold theory we obtain sufficient conditions for the
asymptotic stability of de Sitter solution with standard 4D behavior.
We also prove that there are not late time de Sitter attractors
with 5D-modifications since they are always saddle-like. This fact
correlates with a transient primordial inflation. We present here sufficient conditions on the potential for the stability of the  scalar field-matter scaling solution, the scalar field-dominated solution, and the scalar field-dark radiation scaling solution.  We illustrate our analytical findings using a simple $f$-deviser as a toy model.
All these results are generalizations of our previous results obtained for FRW branes.  

\end{abstract}

\maketitle

\section{Introduction}
Randall-Sundrum braneworlds were first proposed in \cite{Randall1999b, Randall1999c}. In these references was proved that for non-factorizable geometries in five dimensions there exists a single massless bound state confined in a domain wall or three-brane. This bound state is the zero mode of the Kaluza-Klein dimensional reduction and corresponds to the four-dimensional graviton. The Randall-Sundrum brane of type $2$ model,
as an alternative mechanism to the Kaluza-Klein compactifications
\cite{Randall1999b}, have been intensively studied in the last
years, among other reasons, because its appreciable cosmological
impact in the inflationary scenario \cite{Hawkins2001, Huey2001, Huey2002}. The setup of the model start with the
particles of the standard model confined in a four dimensional
hypersurface with positive tension embedded in a 5-dimensional
bulk with negative cosmological constant. It is well-known that
the cosmological field equations on the brane are essentially
different from the standard 4-dimensional cosmology \cite{Binetruy2000, Binetruy2000b, Bowcock2000b}.

Friedmann-Robertson branes with an scalar field trapped on it have been investigated widely in the literature \cite{Maeda2001, Gonzalez-Diaz2000, Majumdar2001h, Nunes2002, Sami2004}. In \cite{Copeland2005} was shown that the potential $V \propto cosech^2(A \phi)$ leads to scaling solution in a RS2 scenario. The dynamics of a scalar field with
constant and exponential potentials was investigated by \cite{Gonzalez2009}. These results were extended
to a wider class of self-interaction potential in \cite{Leyva2009}
using a method proposed in \cite{Fang:2008fw} supporting the idea
that this scenario modifies gravity only at very high energy/short
scales (UV modifications only) having an appreciable impact on
primordial inflation but does not affecting the late-time dynamics
of the Universe unless if the energy density of the matter
trapped in the brane increase at late times
\cite{Garcia-Salcedo2011}. In the reference \cite{Escobar2011} we investigate a scalar field with arbitrary
potential trapped in a RS2 Braneworld. There we present sufficient
conditions for the asymptotic stability of de Sitter solution and  for the stability of scaling
solutions as well as for the stability of the scalar-field dominated
solution extending the results in \cite{Copeland1998} to the higher dimensional framework.
We prove the non-existence of late time attractors with
5D-modifications -a fact that correlates with a transient primordial inflation. Finally, as an example we studied a scalar field with exponential potential, \be
V(\phi)=V_{0}e^{-\chi\phi}+\Lambda\label{exppot}\ee confined in the brane. We proved that for $\chi<0$ the de Sitter solution is asymptotically stable. However, for $\chi>0$ we have proved the de Sitter solution is unstable (of saddle type).  The exponential potential \eqref{exppot}
have been widely investigated in the literature. It was
studied for quintessence models in \cite{Cardenas2003} were it is
considered a negative cosmological constant $\Lambda$. In our case
of interest we assume $\Lambda\geq 0,$ to avoid dealing with
negative values of $y.$ However we can apply our procedure by
permitting negative values for $y$ for the case $\Lambda< 0.$ The
dark energy models with exponential potential and negative
cosmological constant were baptized as Quinstant Cosmologies. They
were investigated in \cite{Leon2009a} using an alternative
compactification scheme. The asymptotic properties of a
cosmological model with a scalar field with exponential potential
have been investigated in the context of the General Relativity by
the authors of \cite{Fang:2008fw, Copeland1998}, and in the context
of the RS2 braneworlds  by \cite{Gonzalez2009, Goheer2003}. In both
cases it was studied of the pure exponential potential ($\Lambda=
0$). Potentials of exponential orders at infinity were studied in
the context of Scalar-tensor theories and conformal $F(R)$
theories by the authors of \cite{Leon2009,leon2010, leon2011a}.

Here we give a step further by considering the natural generalization of FRW, i.e., Bianchi I cosmologies. The results in \cite{Escobar2011} concerning the exponential potential apply here too.

Bianchi I models are the minimal extension of FRW metric to the anisotropic framework. Homogeneous but anisotropic geometries are well-known \cite{misner1973gravitation, peebles1993principles}. Bianchi I, Bianchi III, and Kantowski-Sachs can be a very good representation for the homogeneous but anisotropic universe. They were investigated in the framework of $f(R)$ cosmology from both numerical and analytical viewpoint also incorporating the matter content (see \cite{Leon2011} and the references therein).
The evolution of cosmological braneworld models were investigated, for instance, in \cite{Campos2000, Campos2001, Campos2001a, Campos2003, Fabbri2004a, Niz2008}. In \cite{Campos2001a} it is presented a systematic analysis of FRW, Bianchi I and Bianchi V  metrics in these scenarios. There it is discussed the changes in the structure of the phase space with respect the general-relativistic case. In \cite{Maartens2001} it is studied the dynamics of a BI brane in the presence of inflationary scalar field and it is showed that the high energy effects from extra dimensional gravity removes the anisotropic behavior near the initial singularity which is found in general relativity. However, if an anisotropic stresses are included the model behavior in the vicinity of the initial singularity changes. In \cite{Barrow2002} a BI brane-world model with a pure magnetic field and a perfect fluid with a linear barotropic $\gamma$-law equation of state was studied being found that if $\gamma\leq \frac{4}{3}$ \footnote{If $\gamma > \frac{4}{3}$
the previous fixed point is not the only past attractor possibility. It was argued, in both cases, that chaotic behavior is not possible \cite{Barrow2002, Coley2002}.} the past attractor corresponds to a critical point with non trivial magnetic field, which is also a local source, bringing the anisotropic behavior back in the initial singularity.

In this paper we are interested in investigate the evolution of an
scalar field with arbitrary potential trapped in a
Randall-Sundrum's Braneworld of type 2. We consider an
homogeneous but anisotropic Bianchi I (BI) brane filled with a
mixture of an scalar field with arbitrary potential and a perfect
fluid with equation of state (EoS) parameter $\omega=\gamma-1$
with $ 1\leq\gamma\leq2$. To the potential treatment we use the
method developed in \cite{Fang:2008fw}. We extend previous results in
\cite{Campos2001a, Goheer2003, Gonzalez2009, Leyva2009, Escobar2011} by
considering scalar fields with arbitrary potentials in an
anisotropic background and considering the effect of the
projection of the five-dimensional Weyl tensor onto the
three-brane in the form of a positive Dark Radiation term. In  order to illustrate our analytical results we consider a toy model using the simple
$f$-deviser given by $f(s)=s+\alpha.$ The cosmological implications of such a model are also discussed.

\section{Bianchi cosmology in the brane}
The set up is as follows. Using the Gauss-Codacci equations, relating the four and
five-dimensional spacetimes, we obtain the modified Einstein equations on the brane \cite{Shiromizu2000, Sasaki2000}\footnote{In this section, we follow \cite{Campos2001, Goheer2003} in order to give a brief outline of the considerations that lead to the effective Einstein equations for Bianchi I models (\ref{2.4}-\ref{klein}).}:
\be G_{ab}=-\Lambda_4g_{ab}+\kappa^2T_{ab}+\kappa_{(5)}^4S_{ab}-{\cal E}_{ab}\label{2.3}
\ee
where $g_{ab}$ is the four-dimensional metric on the brane and
$G_{ab}$ is the Einstein tensor, $\kappa$ is the four-dimensional gravitational
constant, and $\Lambda_4$ is the cosmological constant induced in the brane. $S_{ab}$ are quadratic corrections in the matter variables. Finally, ${\cal E}_{ab}$ are corrections coming from the extra dimension. More precisely, ${\cal E}_{ab}$
are the components of the electric part of the Weyl tensor of the bulk (see \cite{Shiromizu2000} and the review \cite{Maartens2004}). Following \cite{Maartens2000c, Campos2001, Goheer2003}, from the energy-momentum tensor conservation equations ($\nabla_aT^a_b=0$) and equations (\ref{2.3}) we get a constraint on $S_{ab}$ and ${\cal E}_{ab}$:
\begin{equation}
\nabla^a\left({\cal E}_{ab}-\kappa_{(5)}^4S_{ab}\right)=0\label{new4}
\end{equation}
In general we can decompose ${\cal E}_{ab}$ with respect to a chosen 4-velocity field $u^a$\cite{Maartens2000c}
as:
\begin{equation}
 {\cal E}_{ab}=-\left(\frac{\kappa_{(5)}}{\kappa}\right)^4\left[{\cal U}(u_a u_b+\frac{1}{3}h_{ab})+{\cal P}_{ab}+2u_{(a}{\cal Q}_{b)} \right]
\end{equation}
where
\begin{equation}
 {\cal P}_{(ab)}={\cal P}_{ab},\;\;\;{\cal P}^a_a=0,\;\;\;{\cal P}_{ab}u^b=0,\;\;\;{\cal Q}_a u^a=0
\end{equation}
here the scalar component ${\cal U}$ is referred as the dark energy density due to it has the same form as the energy-momentum tensor of a radiation perfect fluid. ${\cal Q}_a$ is an spatial vector that corresponds to an effective nonlocal energy flux on the brane and ${\cal P}_{ab}$ is an spatial, symmetry and trace-free tensor which is an effective non local anisotropic stress. Another important point is that the $9$ independent component in the trace-free ${\cal E}_{ab}$ are reduced to $5$ degree of freedom by equation (\ref{new4}) \cite{Maartens2004, Maartens2010}, i.e., the constraint equation (\ref{new4}) provides evolution equations for ${\cal U}$ and ${\cal Q}_a$, but not for ${\cal P}_{ab}$.

Taking into account the effective Einsteins equations (\ref{2.3}), the consequence of having a Bianchi I model\footnote{the metric in the brane is given by $ds^2 =-a_{0}^2 (t)dt^2 + \sum a_{i}(t)(dx^i)^2$} on the brane is \cite{Maartens2001}:
\begin{equation}
 {\cal Q}_a=0 \label{condi12}
\end{equation}
but we do not get any restriction on ${\cal P}_{ab}$ \cite{Campos2001, Goheer2003}. Since there is no way of fixing the dynamics of this tensor we will study the particular case in which:
\begin{equation}
 {\cal P}_{ab}=0,\label{condi13}
\end{equation}
this condition, toghether with (\ref{condi12}) and (\ref{new4}), implies:
\begin{equation}
 D_a{\cal U}=0\Leftrightarrow{\cal U}={\cal U}(t)
\end{equation}

Using the above conditions over ${\cal Q}_a$ and ${\cal P}_{ab}$ (\ref{condi12}-\ref{condi13}), setting the effective cosmological constant in the brane to zero, i.e., $\Lambda_4=0$ \footnote{The induced cosmological constant in the brane can be set to $\Lambda_4=0$ by fine tuning the negative cosmological constant of the $AdS_5$ with the positive brane tension $\lambda>0$  \cite{Brax2003, Maartens2004}.}, the effective Einstein equations (\ref{2.3}) for Bianchi I models (which have zero 3-curvature, i.e., $R^{(3)}=0$)
become:
\be H^{2}= \frac{1}{3}\kappa^{2}\rho_{T}\left(1+ \frac{\rho_{T}}{2\lambda}\right)+ \frac{1}{3}\sigma^2  + \frac{2{\cal U}}{\lambda}\label{2.4}
\ee
\be \dot{H}=- \frac{1}{2}\left(1+ \frac{\rho_T}{\lambda}\right)\left(\dot{\phi}^2 +\gamma\rho_m \right)-\frac{4{\cal U}}{\lambda}-\sigma^2\label{2.5}
\ee
\be \dot{\sigma}=- 3 H\sigma\label{2.5b}
\ee
\be
\dot{{\cal U}}=-4H{\cal U}
\ee
\be \dot{\rho}_m+3H(\rho_m+p_m)=0\label{2.6}
\ee
\be \ddot\phi-3H\dot\phi+\partial_\phi V=0\label{klein},
\ee
where $\sigma$ is the shear term.

It is also convenient to relate $p$ and $\rho$ by
\be p=(\gamma-1)\rho.
\ee

The Dark Radiation term in (\ref{2.4}-\ref{2.5}) evolves as ${\cal U}(t)=\frac{C}{a(t)^4}$ \cite{Maartens2000c}, where $C$ is a constant parameter. An important issue comes from the sign that can take $C$. From the point of view of the brane (brane base formalism) $C$ is just an integrations constant and can take any sign. However it was shown, from the bulk base formalism, that for all possible homogeneous and isotropic solutions on the brane, the bulk spacetime is Schwarzschild-AdS being possible to identify $C$ with the mass of a bulk black hole ($\mu=\frac{2 C}{\lambda}$), and use this to constrain to be positive \cite{Mukohyama2000, Maartens2010, Clifton2012}. For anisotropic models such constraints do not exist, i.e., $C$, and therefore ${\cal U}$, can take any sign \cite{Campos2001, Santos2001e, Goheer2003}. In this first approach for investigating arbitrary potentials, we remain in the case $C>0\implies {\cal U}>0.$ The case of ${\cal U}<0$ is also worthy of investigation and deserves a careful analysis. For  investigating these classes of models, i.e., with ${\cal U}<0,$ it is requiered to consider a new set of variables than in the present paper (more precisely, normalizing with $D=\sqrt{H^2-\frac{2 U}{\lambda}}$ instead of $H$ and reemplacing the analogue to the variable $y$ but the new one $Q=\frac{H}{D}$) \footnote{This allows for the construction of a phase space compact in all the variables but  $s$ defined in eq. \eqref{4}, that runs, in principle, over all the real values.}. Since the resulting system have its  own subtetlies and also have a rich cosmological behavior,  these results would enlarge the present report, so we prefer to address the point in a forthcomming paper \cite{Leon2012}.

\section{Dynamical systems analysis for Bianchi I brane with positive Dark Radiation term}

\subsection{Dynamical variables and dynamical system}

In this section we investigate the cosmological model \eqref{2.4}-\eqref{klein} from the dynamical systems perspective.
First we need to recast the cosmological equations \eqref{2.4}-\eqref{klein} as an autonomous system of first order ordinary differential equations \cite{coley2003dynamical, Tavakol2005B, leon2010, leon2011a}. For this purpose we introduce the normalized variables \bea
x=\frac{\dot{\phi}}{\sqrt{6}H}\qquad y= \frac{V}{3H^2}\qquad \Omega_\lambda = \frac{\rho_{T}^2}{6\lambda H^2}\nonumber \\
\Omega_m =\frac{\rho_m}{3H^2}\qquad \Omega_\sigma =\frac{\sigma}{\sqrt{3}H}\qquad \Omega_U = \frac{2{\cal U}}{\lambda H^2} \label{var22},
\eea
the
additional dynamical (non-compact) variable, $s$, given by \be
s=-\partial_{\phi} \ln\;V(\phi).\label{4} \ee which is a function
of the scalar field and the new time variable $\tau=\ln a(t).$

For the scalar potential treatment, we proceed following the reference \cite{Fang:2008fw, Escobar2011}. Let be defined
the scalar function \be f=\Gamma-1,\qquad \Gamma=
\frac{V^{''}V}{V^{'2}}\label{function}.\ee Since $\Gamma$ is a
function of the scalar field $\Gamma(\phi)$  (see definition
\eqref{function}), also is the variable $s=S(\phi).$ Assuming that
the inverse of $S$ exists, we have  $\phi=S^{-1}(s).$ Thus, one
can obtain de relation $\Gamma=\Gamma(S^{-1}(s))$ and finally the
scalar field potential can be parameterized by a function $f(s).$

Using the Friedmann equation \eqref{2.5} we obtain the following relation between the variables  \eqref{var22}
\be x^2 +y +\Omega_m +\Omega_{\lambda}+\Omega_{\sigma}^2 +\Omega_{U}=1\label{e2.60}
\ee
The restriction \eqref{e2.60} allows to forget about one of the dynamical variables, e.g., $\Omega_U,$  to obtaining a reduced dynamical system. From the condition $0\leq\Omega_{U}\leq1$ we have the following inequality
\be 0\leq x^2 +y +\Omega_m +\Omega_{\lambda}+\Omega_{\sigma}^2 \leq1\label{e2.61}
\ee

Using the variables \eqref{var22}, the field equations \eqref{2.4}-\eqref{klein} and the new time variable we obtain the following autonomous system of ordinary differential equations (ODE):
\begin{align}
 &x '= \sqrt{\frac{3}{2}} s y+ x^3+\left(\Omega_\sigma ^2-2 y+\left(\frac{3 \gamma }{2}-2\right) \Omega_m+\right. \nonumber \\& \left.+\frac{3 ((\gamma -2) \Omega_m-2 y)
   \Omega_\lambda }{x^2+y+\Omega_m}+4 \Omega_\lambda -1\right) x\label{eqx}
\\
&y '=y \left(2 x^2-\sqrt{6} s x+2 \Omega _{\sigma }^2-4 y+3 \gamma  \Omega _m-4 \Omega _m+\right.\nonumber \\ & \left. +\frac{2 \left(4 x^2-2 y+(3 \gamma -2) \Omega
   _m\right) \Omega _{\lambda }}{x^2+y+\Omega _m}+4\right)\label{eqy}
\\
&\Omega_m '=\Omega _m \left(2 x^2-4 y-3 \gamma +(3 \gamma -4) \Omega _m+4\right)+\nonumber\\&+\frac{\Omega _m \left(8
   x^2-4 y+(6 \gamma -4) \Omega _m\right) \Omega _{\lambda }}{x^2+y+\Omega _m}+2 \Omega _m \Omega _{\sigma }^2\label{eqm}
\\
&\Omega_\lambda '=\frac{\left(8 x^2-4 y+(6 \gamma -4) \Omega _m\right) \Omega _{\lambda }^2}{x^2+y+\Omega _m}+\nonumber\\ &+2 \left(x^2-2 y-3 \gamma +\frac{3
   \left((\gamma -2) x^2+y \gamma \right)}{x^2+y+\Omega _m}+2\right) \Omega _{\lambda }+\nonumber\\ &+\left(2 \Omega _{\sigma }^2+(3 \gamma -4)
   \Omega _m\right) \Omega _{\lambda }\label{eqlambda}
\\
&\Omega_\sigma '=\Omega _{\sigma }^3+\left(x^2-2 y+\left(\frac{3 \gamma }{2}-2\right) \Omega _m-1\right) \Omega _{\sigma }+\nonumber\\ & +\frac{\left(4 x^2-2
   y+(3 \gamma -2) \Omega _m\right) \Omega _{\lambda } \Omega _{\sigma }}{x^2+y+\Omega _m}\label{eqsigma}
\\
& s'=-\sqrt{6} s^2 x f(s)\label{eqs}
\end{align}

Using the Friedmann equation \eqref{e2.60} we obtain the following useful relationship
\be \frac{\rho_{T}}{\lambda}=\frac{\Omega_{\lambda}}{\Omega_{m}+x^2 +y}.\label{e2.63}
\ee
From \eqref{e2.63} follows that the region  $\Omega_{m}+x^2 +y=0$ corresponds to cosmological solutions where
$\rho_T \gg \lambda$ (corresponding to the formal limit
$\lambda\rightarrow 0$). Therefore, they are associate to high
energy regions, i.e., to cosmological solutions in a neighborhood
of the initial singularity \footnote{See the references
\cite{Foster1998, Leon2009} for a classical treatment of
cosmological solutions near the initial singularity in FRW cosmologies.}. Due to its
classic nature, our model is not appropriate to describing the
dynamics near the initial singularity, where quantum effects
appear. However, from the mathematical viewpoint, these region
($\Omega_{m}+x^2 +y=0$) is reached asymptotically. In fact, as
some numerical integrations corroborate, there exists an open set
of orbits in the phase interior that tends to the boundary
$\Omega_{m}+x^2 +y=0$ as $\tau\rightarrow-\infty$. Therefore, for
mathematical motivations it is common to attach the boundary
$\Omega_{m}+x^2 +y=0$ to the phase space \footnote{We submitt the reader to the end of section \ref{section3.2} for a discussion of how to dealing with the singular points at the hypersurface $\Omega_{m}+x^2 +y=0.$}.  On the other hand the
points with ($\Omega_\lambda=0$) are associated to the standard
4D behavior ($\rho_T \ll \lambda$ or $\lambda\rightarrow\infty$)
and corresponds to the low energy regime.

From definition \eqref{var22} and from the restriction \eqref{e2.60},
and taking into account the previous statements, it is enough to
investigate to the flow of \eqref{eqx}-\eqref{eqs} defined in the
phase space
\bea \Psi=\{(x,y,\Omega_{m},\Omega_{\lambda},\Omega_{\sigma}): 0\leq x^2 +y +\Omega_m\nonumber\\
+\Omega_{\lambda}+\Omega_{\sigma}^2 \leq1,-1\leq x\leq1,0\leq y\leq1,\nonumber\\
 0\leq\Omega_{m}\leq1, 0\leq\Omega_{\lambda}\leq1,-1\leq\Omega_{\sigma}\leq1\}\times\left\{s\in\mathbb{R}\right\}\label{reg}.
\eea

\begin{table*}\caption[crit2]{Existence conditions for the critical point of the system  \eqref{eqx}-\eqref{eqs}. We use the notations $s_c$ for an arbitrary real s-value and $s^*$ for an s-value such that $f(s^*)=0.$}
\begin{center}\begin{tabular}{@{\hspace{4pt}}c@{\hspace{10pt}}c@{\hspace{10pt}}c@{\hspace{10pt}}c@{\hspace{10pt}}c@{\hspace{10pt}}c@{\hspace{10pt}}c@{\hspace{10pt}}c}
\hline
\hline\\[-0.3cm]
$P_i$ &$x$& $y$& $\Omega_m$& $\Omega_\lambda$& $\Omega_{\sigma}$& $s $& Existence \\[0.1cm]
\hline\\[-0.2cm]
$P_{1}$& 0 & 0 & 1& 0 & 0& $s_c$ & $s_c\in\mathbb{R}$ \\[0.2cm]
$P_{2}^\pm$& $\pm 1$& 0&  0 &  0&  0 &  0&  always \\[0.2cm]
$P_{3}^\pm$& $\pm 1$ & 0 & 0& 0& 0& $s^*$ & always \\[0.2cm]
$P_{4}$& $\frac{\sqrt{\frac{3}{2}} \gamma }{s^*}$ & $-\frac{3 (\gamma -2) \gamma }{2 {s^*}^2}$ & $1-\frac{3 \gamma
   }{{s^*}^2}$ & 0 & 0 & $s^*$ & ${s^*}^2 \geq 3\gamma$\\[0.2cm]
$P_{5}$& $\frac{s^*}{\sqrt{6}}$ & $1-\frac{{s^*}^2}{6}$ & 0 & 0 & 0& $s^*$& $s^{*2}\leq 6$\\[0.2cm]
$P^\pm_{6}$& $\frac{\sqrt{6}}{s^*}$ & 0 & 0 & 0 & $\pm\sqrt{1-\frac{6}{{s^*}^2}}$ & $s^*$ & $s^{*2}\geq 6$\\[0.2cm]
$P_{7}$& $\frac{2 \sqrt{\frac{2}{3}}}{s^*}$ & $\frac{4}{3 {s^*}^2}$ & 0 & 0 & 0& $s^*$& $s^{*2}\geq 2$\\[0.2cm]
$P_{8}^\pm$& $x\in[-1,1]$& $0$& $0$& $0$& $\pm\sqrt{1-x^2}$& $0$& $x\neq0$\\[0.2cm]
$P_{9}^\pm$& $x\in[-1,1]$& $0$& $0$& $0$& $\pm \sqrt{1-x^2}$& $s^*$& $x\neq0$\\[0.2cm]
$P_{10}$& $0$& $1-\Omega_\lambda$& $0$& $\Omega_\lambda\in[0,1[$& $0$& $0$& $\Omega_\lambda\in[0,1[$\\[0.2cm]
$P_{11}$& $0$& $0$& $0$& $1$& $0$& $s_c\in\mathbb{R}$ & always \\[0.4cm]\hline \hline
\end{tabular}\end{center}\label{tab1}
\end{table*}

\begin{table*}\caption{Some basic observables $\omega_{\phi}$, $\Omega_{\phi}$ and $q$ for the critical point of the system  \eqref{eqx}-\eqref{eqs}. We use the notations $s_c$ for an arbitrary real s-value and $s^*$ for an s-value such that $f(s^*)=0.$}
\begin{center}\begin{tabular}{@{\hspace{4pt}}c@{\hspace{10pt}}c@{\hspace{10pt}}c@{\hspace{10pt}}c}
\hline
\hline\\[-0.3cm]
$P_i$  $\omega_\phi$& $\Omega_\phi$& $q$\\[0.1cm]
\hline\\[-0.2cm]
$P_{1}$& {undefined} & $0$& $\frac{3\gamma}{2}-1$\\[0.2cm]
$P_{2}^\pm$& $1$& $1$& $2$\\[0.2cm]
$P_{3}^\pm$& $1$& $1$& $2$\\[0.2cm]
$P_{4}$&$\gamma-1$& $\frac{3\gamma}{s^{*2}}$& $\frac{3\gamma}{2}-1$\\[0.2cm]
$P_{5}$& $\frac{s^{*2}}{3}-1$& $1$& $\frac{s^{*2}}{2}-1$\\[0.2cm]
$P^\pm_{6}$& $1$& $\frac{6}{s^{*2}}$& $\frac{18}{s^{*2}}-1$\\[0.2cm]
$P_{7}$& $\frac{1}{3}$& $\frac{4}{s^{*2}}$& $\frac{8}{s^{*2}}-1$\\[0.2cm]
$P_{8}^\pm$&$1$& $x^2$& $3x^2-1$\\[0.2cm]
$P_{9}^\pm$& $1$& $x^2$& $3x^2-1$\\[0.2cm]
$P_{10}$&  $-1$& $1-\Omega_\lambda$& $-1$\\[0.2cm]
$P_{11}$&  undefined & 0 & undefined \\[0.4cm]\hline \hline
\end{tabular}\end{center}\label{tab1b}
\end{table*}

\subsection{Phase space analysis }\label{section3.2}

The critical points of the system \eqref{eqx}-\eqref{eqs} are summarized in Table \ref{tab1} where $s^*$ corresponding $f(s^*)=0$.
All the points showed in table \ref{tab1} but $P^{\pm}_8$, $P^{\pm}_9$ for $x=0$, and  $P_{10}$ when $y=0$ always satisfy the condition $\Omega_{m}+x^2 +y\neq0.$ The points where this condition do not holds will be excluded from our analysis, since they represents solutions near the initial singularity. As we commented in the last section, our model is not applicable near this singularity.

It will be helpful to have the important observational parameters in terms of the state variables. The dimensionless  scalar field energy density parameter $\Omega_\phi=\frac{\rho_{\phi}}{3H^2}$, the equation of state parameter $\omega_{\phi}= \frac{p_{\phi}}{\rho_{\phi}}$, and deceleration parameter $q=-\left(1+\frac{\dot{H}}{H^2}\right).$ They read:
\be \Omega_\phi=x^2+y, \qquad \omega_{\phi}=\frac{x^2-y}{x^2+y},
\ee
\be q=-1+3x^2+\frac{3\gamma}{2}\Omega_m+\frac{3\left(2x^2+\gamma\Omega_m\right)}{x^2+y+\Omega_m}
\ee

In the table \ref{tab1} are displayed the existence conditions for the critical point of the system  \eqref{eqx}-\eqref{eqs} whereas in the table \ref{tab1b} are shown the values of the cosmological parameters  $\omega_{\phi}$, $\Omega_{\phi}$ and $q$.

\begin{table*}[tbp]\caption[crit2]{Eigenvalues for the critical points in Table \ref{tab1}. We use the notation $\beta_\pm=\frac{3}{4} \left(\gamma -2\pm\sqrt{(2-\gamma ) \left(\frac{24 \gamma
   ^2}{{s^*}^2}-9 \gamma +2\right)}\right).$}
\begin{center}\begin{tabular}{@{\hspace{4pt}}c@{\hspace{14pt}}c@{\hspace{14pt}}c}
\hline
\hline\\[-0.3cm]
$P_i$ & Eigenvalues\\[0.1cm]
\hline\\[-0.2cm]
$P_{1}$& $\left\{-3 \gamma ,3 \gamma ,\frac{3 (\gamma -2)}{2},\frac{3 (\gamma -2)}{2},3 \gamma -4,0\right\}$\\[0.2cm]
$P_{2}^{\pm}$& $\{-6,6,2,0,0,6-3 \gamma \}$\\[0.2cm]
$P_{3}^{\pm}$& $\left\{-6,2,0,0,6-3 \gamma ,6\mp\sqrt{6} s^*, \mp \sqrt{6} {s^*}^2 f'\left(s^*\right)\right\}$\\[0.2cm]
$P_{4}$& $\left\{3 \gamma -4,\frac{3 (\gamma -2)}{2},-3 \gamma,\beta_+,\beta_-,-3 \gamma  s^* f'\left(s^*\right)\right\}$\\[0.2cm]
$P_{5}$& $\left\{\frac{1}{2} \left({s^*}^2-6\right),\frac{1}{2}
   \left({s^*}^2-6\right),{s^*}^2-4,{s^*}^2-3 \gamma ,-{s^*}^2,-{s^*}^3
   f'\left(s^*\right)\right\}$\\[0.2cm]
$P^\pm _{6}$& $\left\{2,0,0,6-3 \gamma ,-6,-6 s^* f'\left(s^*\right)\right\}$\\[0.2cm]
$P_{7}$& $\left\{-1,4-3 \gamma ,-4,-\frac{1}{2} \left(1+\sqrt{\frac{64}{{s^*}^2}-15}\right),-\frac{1}{2} \left(1-\sqrt{\frac{64}{{s^*}^2}-15}\right),-4 s^* f'\left(s^*\right)\right\}$\\[0.2cm]
$P_{8}^{\pm}$& $\{6,2,0,0,-6,6-3 \gamma \}$\\[0.2cm]
$P_{9}$& $\left\{2,0,-6,6-3 \gamma ,6-\sqrt{6} x s^*,-\sqrt{6} x {s^*}^2 f'\left(s^*\right)\right\}$\\[0.2cm]
$P_{10}$& $\{-4,0,0,-3,-3,-3 \gamma \}$
\\[0.2cm]
$P_{11}$& $\{0,3 (\gamma -1),3 (\gamma -1),3 \gamma ,6 \gamma ,6 \gamma -4\}$\\[0.4cm]\hline \hline
\end{tabular}\end{center}\label{autovalores1}
\end{table*}

Now, let us comment on the stability and physical interpretation of the critical points/curves.

The point $P_1$ represents a matter-dominated solution
($\Omega_m=1$). Although it is non-hyperbolic, it behaves like
a saddle point in the phase space of the RS model, since
they have both nonempty stable and unstable manifolds (see the table \ref{autovalores1}).

The critical points $P^\pm_2$ are solutions dominated by the
kinetic energy of the scalar field and they represent solutions
with an standard behavior ($\Omega_{\lambda}=0$). This critical
points are nonhyperbolic. However, they behave as saddle-like
points in the phase space because of the instability in the
eigendirection associated with two positive eigenvalues and the
stability of an eigendirection associated to a negative
eigenvalue.

The critical point $P^\pm_3$ are solutions dominated by the
kinetic energy of the scalar field. They are non-hyperbolic, however they behave as saddle points since they have both nonempty stable and unstable manifolds (see the table \ref{autovalores1}). Thus, they represent transient states
in the evolution of the universe.

The critical point $P_4$ is nonhyperbolic for $\gamma=2,4/3$, $s^*=\pm\sqrt{3\gamma}$ and $f'(s^*)=0$ corresponding to scalar field-matter scaling solutions ($\Omega_{\phi}\sim \Omega_m$).

$P_4$ is a stable node in the cases
$0<\gamma \leq \frac{2}{9},s^*<-\sqrt{3\gamma },f'\left(s^*\right)<0$ or
$\frac{2}{9}<\gamma <\frac{4}{3},-\frac{2 \sqrt{6} \gamma }{\sqrt{9 \gamma -2}}\leq s^*<-\sqrt{3\gamma },
   f'\left(s^*\right)<0,$ or $0<\gamma \leq \frac{2}{9},s^*>\sqrt{3\gamma }, f'\left(s^*\right)>0,$ or $\frac{2}{9}<\gamma <\frac{4}{3},\sqrt{3\gamma }<s^*\leq \frac{2 \sqrt{6} \gamma }{\sqrt{9 \gamma -2}},
   f'\left(s^*\right)>0.$

It is a spiral stable point for either $\frac{2}{9}<\gamma <\frac{4}{3},s^*<-\frac{2 \sqrt{6} \gamma }{\sqrt{9 \gamma -2}},f'\left(s^*\right)<0$ or $\frac{2}{9}<\gamma <\frac{4}{3},s^*>\frac{2 \sqrt{6} \gamma }{\sqrt{9 \gamma -2}},f'\left(s^*\right)>0.$

In summary, it is stable for either $0<\gamma <\frac{4}{3},s^*<-\sqrt{3\gamma },f'\left(s^*\right)<0$ or $0<\gamma <\frac{4}{3},s^*>\sqrt{3\gamma },f'\left(s^*\right)>0.$ Otherwise, it is a saddle point.

The critical point $P_5$ represents a scalar-field dominated solution ($\Omega_\phi=1$) that is not hyperbolic for $s^*\in\left\{0,\pm\sqrt{6},\pm\sqrt{3\gamma},2\right\}$ or $f'(s^*)=0$.

In the hyperbolic case, $P_5$ is a stable node for either $0<\gamma \leq \frac{4}{3},-\sqrt{3 \gamma }<s^*<0,f'\left(s^*\right)<0$ or $\frac{4}{3}<\gamma \leq 2,-2<s^*<0,f'\left(s^*\right)<0$ or $0<\gamma \leq \frac{4}{3},0<s^*<\sqrt{3 \gamma },f'\left(s^*\right)>0$ or $\frac{4}{3}<\gamma \leq 2,0<s^*<2,f'\left(s^*\right)>0;$ otherwise, it is a saddle point.

The critical points $P^\pm _6$ are nonhyperbolic and correspond to scalar field-anisotropic scaling solutions
($\Omega_{\phi}\sim \Omega_{\sigma}$). For $\frac{6}{s^{*2}}\ll1$ the anisotropic term in the Friedmann equation (\ref{2.5}) dominates the cosmological dynamic ($\Omega_{\sigma}=\pm1$). However, they behave as
saddle-like points since they have both nonempty stable and unstable manifolds (see the table \ref{autovalores1}).

The critical point $P_7$ is nonhyperbolic for $\gamma=4/3$, $s^*=\pm2$ and $f'(s^*)=0$ and corresponds to scalar field-dark radiation scaling solutions ($\Omega_{\phi}\sim \Omega_{U}$).

It is a stable node for either $\frac{4}{3}<\gamma \leq 2,-\frac{8}{\sqrt{15}}\leq s^*<-2,f'\left(s^*\right)<0$ or $\frac{4}{3}<\gamma \leq 2,2<s^*\leq \frac{8}{\sqrt{15}},f'\left(s^*\right)>0.$

It is a spiral point for either $\frac{4}{3}<\gamma \leq 2,s^*<-\frac{8}{\sqrt{15}},f'\left(s^*\right)<0$ or $\frac{4}{3}<\gamma \leq 2,s^*>\frac{8}{\sqrt{15}},  f'\left(s^*\right)>0.$

In summary, it is an attractor for either $\frac{4}{3}<\gamma \leq 2,s^*<-2,  f'\left(s^*\right)<0$ or $\frac{4}{3}<\gamma \leq 2,s^*>2,  f'\left(s^*\right)>0;$ it is a saddle otherwise.

The circles of critical points $P^\pm_8$ and $P^\pm_9$ are nonhyperbolic and corresponding to scalar field-anisotropic scaling solutions ($\Omega_{\phi}\sim \Omega_{\sigma}$). Both solutions represents transient states in the evolution of the universe. When $x\rightarrow0$ the anisotropic term in the Friedmann equation (\ref{2.5}) dominates the cosmological dynamics ($\Omega_{\sigma}=\pm1$). In this limit the corresponding cosmological are in a vicinity of the initial singularity.

The line of critical points $P_{10}$ is nonhyperbolic. They
represent solutions with 5D-corrections for
$\Omega_{\lambda}\neq0$ whereas for $\Omega_{\lambda}=0$ it represents an standard 4D cosmological solution. From the relationship between $y$ and
$\Omega_{\lambda}$ follows that this solution is dominated by the
potential energy of the scalar field $\rho_T= V(\phi);$ that is,
it is de Sitter-like  solution ($\omega_\phi=- 1$). In this case
the Friedmann equation can be expressed as \be 3H^2 =V\left(1+
\frac{V}{2\lambda}\right)\label{P10.1} \ee In the early universe,
where $\lambda\ll V,$ the expansion rate of the universe for the
RS model differs from the general relativity predictions
\be\frac{H_{RS}}{H_{GR}}= \sqrt{\frac{V}{2\lambda}}\label{P10.2} \ee
Due the importance of de Sitter solutions in the cosmological context, in the section
\ref{CMP10} we  calculate explicitly  their center manifold. Due the physical differences between solutions with standard 4D and non-standard 5D behaviors, we consider the cases $\Omega_\lambda=0$ and $\Omega_\lambda\neq 0$ in separated subsections.

Summarizing, the possible late-time stable solutions are
\begin{itemize}
\item The scalar field-matter scaling solution ($P_4$) provided $0<\gamma <\frac{4}{3},s^*<-\sqrt{3\gamma },f'\left(s^*\right)<0$ or $0<\gamma <\frac{4}{3},s^*>\sqrt{3\gamma },f'\left(s^*\right)>0.$
\item The scalar field-dominated solution ($P_5$) provided $0<\gamma \leq \frac{4}{3},-\sqrt{3 \gamma }<s^*<0,f'\left(s^*\right)<0$ or $\frac{4}{3}<\gamma \leq 2,-2<s^*<0,f'\left(s^*\right)<0$ or $0<\gamma \leq \frac{4}{3},0<s^*<\sqrt{3 \gamma },f'\left(s^*\right)>0$ or $\frac{4}{3}<\gamma \leq 2,0<s^*<2,f'\left(s^*\right)>0.$
\item The scalar field-dark radiation scaling solution ($P_7$) provided $\frac{4}{3}<\gamma \leq 2,s^*<-2,  f'\left(s^*\right)<0$ or $\frac{4}{3}<\gamma \leq 2,s^*>2,  f'\left(s^*\right)>0.$
\item The de Sitter solution $P_{10}$ with $\Omega_\lambda=0$ provided $f(0)$ is a real (finite) positive number, i.e., $f(0)>0.$
\end{itemize}

It is worth noticing that there  are fractional terms with $x^2+y+\Omega_m$ in the denominator, in the equations \eqref{eqx}-\eqref{eqs}. To finish this section we want to discuss the case when the denominator get close to zero and the effects this may have for the system.
First than all, the only singular solution located in the hypersurface $x^2+y+\Omega_m=0$ is the fixed point $P_{11}.$
In order to investigate the dynamics of  \eqref{eqx}-\eqref{eqs} near $P_{11}$ we introduce the local coordinates
\begin{align}
&\left\{x,y,\Omega_\lambda-1, \Omega_\sigma,\Omega_m, s-s_c\right\}= \nonumber\\ &\epsilon \left\{\widetilde{x},\widetilde{y},\widetilde{\Omega_\lambda}, \widetilde{\Omega_\sigma},\widetilde{\Omega_m},\widetilde{s}\right\}+{\cal O}(\epsilon)^2,
\end{align}
where $\epsilon$ is a constant satisfying $\epsilon\ll 1.$

\begin{align}
&\widetilde{x}'=\sqrt{\frac{3}{2}} s_c \widetilde{y}+\widetilde{x} \left(\frac{3 \gamma  r}{1+ r}-3\right),\nonumber\\
&\widetilde{y}'=\frac{6 \gamma  \widetilde{y}  r}{1+r},\nonumber\\
&\widetilde{\Omega_\lambda
   }'=\frac{3 \gamma
  r \left(\widetilde{y}+ \widetilde{\Omega_m}+2 \widetilde{\Omega_\lambda
   }\right)}{1+r}-4 \left(\widetilde{y}+ \widetilde{\Omega_m}+\widetilde{\Omega_\lambda
   }\right),\nonumber\\
&\widetilde{\Omega_\sigma }'=3 \widetilde{\Omega_\sigma } \left(\frac{\gamma   r}{1+r}-1\right),\nonumber\\
& \widetilde{\Omega_m}'=\frac{3 \gamma   r \left( \widetilde{\Omega_m}-\widetilde{y}\right)}{1+r},\nonumber\\
&\widetilde{s}'=-\sqrt{6} {s_c}^2 \widetilde{x} f(s_c)\label{systemP11}
\end{align} where $$r=\frac{\widetilde{\Omega_m}}{\widetilde{y}}.$$
From the above equations it is deduced the equation
$r'=-3\gamma r.$ Since $\gamma>0,$ follows that $r$ goes to zero (resp. to infinity) at an exponential rate as $\tau\rightarrow \infty$ (resp. $\tau\rightarrow -\infty$).
Thus, in the limit $\tau\rightarrow -\infty,$ $r/(1+r)\rightarrow 1,$ and the system \eqref{systemP11} has the asymptotic structure
\begin{align}
&\widetilde{x}'=3(\gamma-1)\widetilde{x}+\sqrt{\frac{3}{2}} s_c\widetilde{y},\nonumber\\
&\widetilde{y}'=6\gamma \widetilde{y},\nonumber\\
&\widetilde{\Omega_\lambda}'=2(3\gamma-2)\widetilde{\Omega_\lambda}+(3\gamma-4)(\widetilde{y}+\widetilde{\Omega_m}), \nonumber\\
&\widetilde{\Omega_\sigma}'=3(\gamma-1)\widetilde{\Omega_\sigma},\nonumber\\
&\widetilde{\Omega_m}'=3\gamma\left(\widetilde{\Omega_m}-\widetilde{y}\right),\nonumber\\
&\widetilde{s}'=-\sqrt{6}{s_c}^2\widetilde{x}f(s_c)\label{systemP11past}
\end{align}
The system \eqref{systemP11past} admits the exact solution passing by $(\widetilde{x}_0, \widetilde{y}_0,\widetilde{{\Omega_\lambda}}_0,\widetilde{{\Omega_\sigma}}_0, \widetilde{{\Omega_m}}_0,\widetilde{s}_0)$ at $\tau=0$ given by
\begin{align}
&\widetilde{x}(\tau )= \frac{e^{3 (\gamma -1) \tau } \left(\sqrt{6} s_c \widetilde{y}_0 \left(e^{3 (\gamma +1) \tau }-1\right)+6
   \widetilde{x}_0 (\gamma +1)\right)}{6 (\gamma +1)},\nonumber\\
&\widetilde{y}(\tau )= \widetilde{y}_0 e^{6 \gamma  \tau },\nonumber\\
&\widetilde{\Omega_\lambda} (\tau )= e^{2 (3
   \gamma -2) \tau } (\widetilde{y}_0+\widetilde{\Omega_m}_0+\widetilde{\Omega_\lambda}_0)-(\widetilde{y}_0+\widetilde{\Omega_m}_0) e^{3 \gamma  \tau
   },\nonumber\\
&\widetilde{\Omega_m}(\tau )= e^{3 \gamma  \tau } \left(\widetilde{y}_0 \left(-e^{3 \gamma  \tau }\right)+\widetilde{y}_0+\widetilde{\Omega_m}_0\right),\nonumber
\end{align}
\begin{widetext}
\begin{align}
&\widetilde{s}(\tau )= \footnotesize{\frac{6 \widetilde{s}_0 \gamma  \left(\gamma ^2-1\right)-s_c^2 f(s_c) \left(s_c \widetilde{y}_0
   \left((\gamma -1) e^{6 \gamma  \tau }-2 \gamma  e^{3 (\gamma -1) \tau }+\gamma +1\right)+2 \sqrt{6} \widetilde{x}_0 \gamma  (\gamma
   +1) \left(e^{3 (\gamma -1) \tau }-1\right)\right)}{6 \gamma  \left(\gamma ^2-1\right)}},\nonumber\\
&\widetilde{\Omega_\sigma} (\tau )= \widetilde{\Omega_\sigma}_0 e^{3 (\gamma -1) \tau }
\end{align}
\end{widetext}
The eigenvalues of the linearization around the fixed points $P_{11},$ given by \eqref{systemP11past}, are displayed in the last row of table  \ref{autovalores1}. They are calculated under the hypothesis $y\ll \Omega_m$ as $y,\Omega_m\rightarrow 0$ (which is consistent with the limit $r\rightarrow \infty$ as $\tau\rightarrow -\infty$).
From the analysis of the system \eqref{systemP11past} follows that the singular point $P_{11}$ represents a Big-Bang singularity. It is non-hyperbolic with a 5D unstable manifold for $1<\gamma<2.$ In this case it is a local source as numerical simulations suggest. For $0<\gamma<1$ it behaves as a saddle point.
Since the limit $\Omega_\lambda\rightarrow 1$ corresponds to the domain where the brane corrections are important, and this regime is associated with the past dynamics, we conclude that the linear approximation \eqref{systemP11} is not valid as $\tau\rightarrow +\infty.$

\section{Dynamics of the center manifold of the de Sitter solutions}\label{CMP10}

In this section we will study the stability of the center manifold of the the de Sitter solutions for $\Omega_{\lambda}= 0$  and for $0<\Omega_{\lambda}<1.$ For this purpose we can use the Center Manifold Theory (see the reviews \cite{leon2010, leon2011a}).

\subsection{Case $\Omega_\lambda=0$}\label{A.1}

The solution $P_{10}$ with $\Omega_{\lambda}= 0$ can be a
candidate to be a late time de Sitter attractor without
5D-corrections. To analyze their stability
we carry out a detailed stability study using the Center
Manifold Theory.

Let us assume that $f(0)\in\mathbb{R}\setminus \{0\}.$
To prepare the system for center manifold calculations we introduce the new variables \bea && u_1= s,\,u_2=\Omega_\lambda,\,v_1=-1+y+\Omega_m+\Omega_\lambda,\,v_2=\Omega_\sigma,\nonumber\\&& v_3=x-\frac{s}{\sqrt{6}},\,v_4=\Omega_m\label{vvar2}.\eea
Then, we Taylor-expand the system $u_1', u_2', v_1', v_2', v_3', v_4'$ in a neighborhood of the origin with error of order $\mathcal{O}(4).$

Accordingly  to the Center Manifold theorem, the local center
manifold of the origin this vector field is given by the graph \bea &
W_{\text{loc}}^c(\mathbf{0})=\left\{(u_1,u_2,v_1,v_2,v_3,v_4):
v_i=f_i(x_1,x_2), \right. \nonumber \\ & \left.  i=1\ldots 4, \mathbf{f}(\mathbf{0})=\mathbf{0}, \mathbf{Df}(\mathbf{0})=\mathbf{0},
u_1^2+u_2^2<\delta\right\}\label{func}\eea  for $\delta>0$ an
small enough real value.

Deriving each one of the functions in (\ref{func}) with respect
$\tau$ and substituting and substituting the vector field $u_1', u_2', v_1', v_2', v_3', v_4'$ one can obtain a system of quasi-lineal partial
differential equations that the functions $f_i$ must satisfy.

Solving this system using Taylor series up to an error term $\mathcal{O}(4)$ we obtain
\bea & v_1=\frac{u_1^2 u_2}{3}-\frac{u_1^2}{6},\, v_2=0,\nonumber\\ & v_3=\frac{u_1^3 f(0)}{3 \sqrt{6}}-\frac{u_1
   u_2}{\sqrt{6}},\,v_4=0.\label{v1v3}
\eea

Thus, the dynamics on the center manifold,  is given by
\bea
u_1'= -f(0) u_1^3 +\mathcal{O}(4)\label{equ1s},\\
u_2'=-u_1^2 u_2 +\mathcal{O}(4)\label{equ2s}.
\eea which is the same given by (36)-(37) in \cite{Escobar2011} with the identifications $u_1\equiv x_1,u_2\equiv x_2.$

Using the same arguments as in \cite{Escobar2011} we obtain that the origin is asymptotic stable for initial conditions in a
vicinity of the origin whenever $f(0)>0.$

From the asymptotic
stability of the origin of (\ref{equ1s})-(\ref{equ2s}) for the
above choice of sign for $f(0),$ follows that the center manifold
of $P_{10}$ (for $\Omega_\lambda=0$) is locally asymptotic stable, and hence, the solution
$P_{10}$ of the system (\ref{eqx})-(\ref{eqs}) also is.

Therefore,
$P_{10}$ with $\Omega_\lambda=0$ and $f(0)>0$ corresponds to a late time de Sitter
attractor. This result for RS-2 brane cosmology is in a perfect
agreement with the standard 4-dimensional TGR framework.

\subsection{Case $\Omega_\lambda\neq 0$}\label{A.2}

In this section we investigate the stability of the curve of
critical points $P_{10}$ for $0<\Omega_{\lambda}<1.$ This solution
corresponds to a de Sitter solution with 5D-corrections.

According
to the RS model this solution cannot behave like a late time
attractor since 5D-corrections are typical of the high energy regime
(early universe) and not to the low energy regime (late universe). If we can
prove that this solution is of saddle type, this
behavior can be correlated with a transient inflationary stage for the universe. In
order to verify our claim we appeal to the Center Manifold Theory.

Let us consider an arbitrary critical point with coordinates
$(x=0, y=1-u_c,\Omega_\lambda=u_c,\Omega_\sigma=0,\Omega_m=0, s=0)$ located at $P_{10}.$

In order to prepare the system (\ref{eqx})-(\ref{eqs}) for the
application of the Center Manifold Theorem we introduce the
coordinate change \bea & u_1=s,\, u_2=-u_c (y+2 \Omega_m)-(u_c-1) \Omega
   _{\lambda },\nonumber\\ & v_1=u_c \left(y+\Omega_m+\Omega _{\lambda
   }-1\right),\, v_2=\Omega_\sigma,\nonumber\\ &  v_3=\frac{s
   (u_c-1)}{\sqrt{6}}+x,\,v_4=u_c
   \Omega_m\eea Then, we Taylor expand the system $u_1', u_2', v_1', v_2', v_3', v_4'$ in a neighborhood
   of the origin with error of order $\mathcal{O}(4).$

Accordingly  to the Center Manifold theorem, the local center
manifold of the origin for the resulting vector field is given by
the graph: \bea &
W_{\text{loc}}^c(\mathbf{0})=\left\{(u_1,u_2,v_1,v_2,v_3,v_4):
v_i=f_i(x_1,x_2), \right. \nonumber \\ & \left.  i=1\ldots 4, \mathbf{f}(\mathbf{0})=\mathbf{0}, \mathbf{Df}(\mathbf{0})=\mathbf{0},
u_1^2+u_2^2<\delta\right\}\label{func2}\eea  for $\delta>0$ an
small enough real value.

Deriving each one of the functions in (\ref{func2}) with respect
$\tau$ and substituting the vector field $u_1', u_2', v_1', v_2', v_3', v_4'$ one can obtain a system of quasi-lineal partial
differential equations that the functions $f_i$ must satisfy.

Solving this system using Taylor series up to an error term $\mathcal{O}(4)$ we obtain
\bea & v_1=-\frac{1}{6} u_1^2 (u_c-1) u_c (2
   u_2+u_c-1),\, v_2=0,\nonumber\\ & v_3=\frac{u_1 \left(u_1^2 (u_c-1)^2 f(0)-3
   u_2\right)}{3 \sqrt{6}},\,v_4=0.\label{p10v1v3}
\eea

Thus, the dynamics on the center manifold,  is given by

\begin{align}
& u_1'= u_1^3 (u_c-1) f(0)+\mathcal{O}(4)\label{p10equ1s},\\
& u_2'=-u_1^2 \left((u_c (3 u_c-4)+1)
   u_2+(u_c-1)^2 u_c\right)+\mathcal{O}(4)\label{p10equ2s}.
\end{align} which is the same as (41), (42) in \cite{Escobar2011} under the variables re-scaling
$$u_1\rightarrow \frac{(1-u_c) u_1}{\sqrt{6}},\, u_2\rightarrow -u_2.$$

Thus, using the same arguments as in \cite{Escobar2011}, the
origin of coordinates is locally asymptotically unstable (of
saddle type) irrespective the sign of $f(0)$. Henceforth, the
center manifold of $P_{10}$ is locally asymptotic unstable (saddle
type) for $f(0)\neq0$.

The physical interpretation
of this result is that there are not late time attractors with
5D-modifications. This type of corrections are characteristic of
the early universe. In this sense the solution $P_{10}$ with $0<\Omega_{\lambda}<1.$
is associated to the primordial inflation.

As in \cite{Escobar2011} for $P_{10}$ with $0\leq\Omega_{\lambda}<1,$ we assume that $f(0)\in\mathbb{R}\setminus \{0\}.$ Otherwise it is required to include higher order terms in the Taylor expansion,
increasing the numerical complexity.

\section{A toy model}
The objective of this section is to illustrate our analytical
results for the following toy model. Let us consider a simple
$f$-deviser given by $f(s)=s+\alpha.$ This function corresponds to
the potential given in the implicit form
\begin{align}
&\phi(s)=\frac{1}{s\alpha}+\ln\left[\left(\frac{s}{s+\alpha}\right)^{\frac{1}{\alpha^2}}\right],\nonumber\\
& V(s)=V_0 \left(\frac{s}{s+\alpha}\right)^{\frac{1}{\alpha}}.
\end{align} or alternatively
\begin{equation}
V(\phi)=V_0 \left(-\frac{1}{W\left(-e^{-\phi  \alpha
^2-1}\right)}\right)^{\frac{1}{\alpha }}
\end{equation} where
$W(z)$ is the special function `ProductLog' that gives the
principal solution for $w$ in $z=we^w.$ $V(\phi)$ is defined for
$\phi\geq 0.$ For $\alpha<0,$ $V$ is a monotonic decreasing
function taking values in the range $[0,V_0];$ for $\alpha>0,$ $V$
is a monotonic increasing function taking values in the range
$[V_0,\infty).$

\begin{figure}
\begin{center}
\includegraphics[height=2in,width=3in]{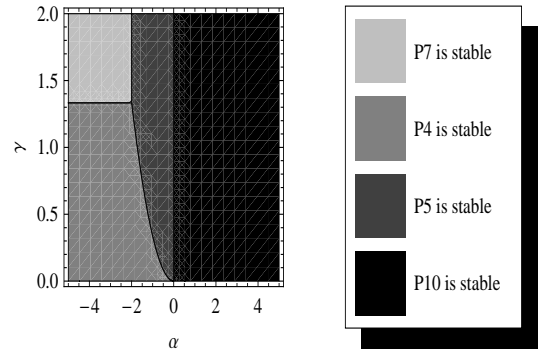}
\caption{\label{parameter} Parameter space diagram showing the
sufficient conditions for the existence of late-time attractors
for the input function $f(s)=s+\alpha.$}
\end{center}
\end{figure}

 The function $f$ satisfies $s^*=-\alpha,$ $f'(s)\equiv 1,$
$f(0)=\alpha$ and $s^*f'(s^*)=s^*=-\alpha.$ The sufficient
conditions for the existence of late-time attractors are fulfilled
easily. They read (see the figure \ref{parameter}):

\begin{itemize}
\item The scalar field-matter scaling solution ($P_4$) is a  late-time attractor provided  $0<\gamma <\frac{4}{3},\alpha<-\sqrt{3\gamma }.$
\item The scalar field-dominated solution ($P_5$) is a late-time attractor provided  $0<\gamma \leq \frac{4}{3},-\sqrt{3 \gamma}<\alpha<0$ or $\frac{4}{3}<\gamma \leq 2,-2<\alpha<0.$
\item The scalar field-dark radiation scaling solution ($P_7$) is a late-time attractor provided  $\frac{4}{3}<\gamma \leq 2,\alpha<-2.$
\item The de Sitter solution $P_{10}$ with $\Omega_\lambda=0$ is the late-time attractor provided $\alpha>0.$
\end{itemize}

\begin{figure}
\begin{center}
\includegraphics[height=2.5in,width=2.5in]{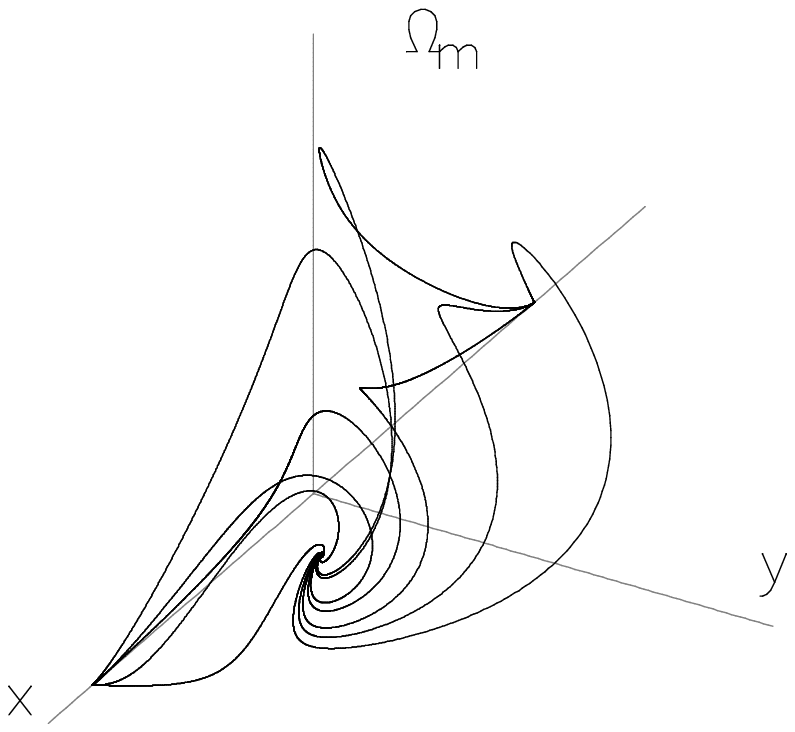}
\begin{center} (a) \end{center}
\includegraphics[height=2.5in,width=2.5in]{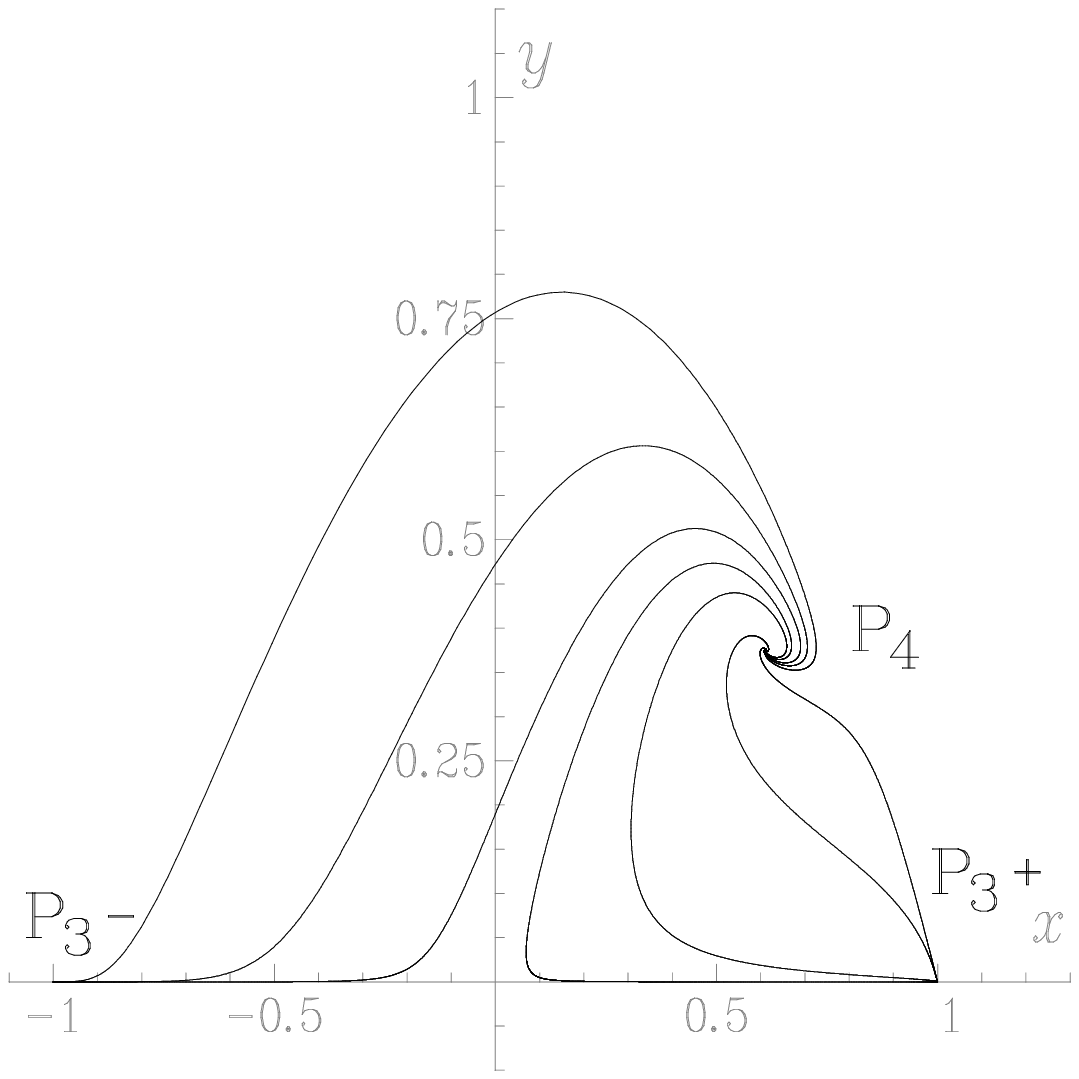}
\begin{center} (b) \end{center}
\caption{\label{P4attractora} (a) Some orbits in the invariant set $\Omega_\lambda=\Omega_\sigma=0, s=2$  of the system
 \eqref{eqx}-\eqref{eqs} for the input function $f(s)=s+\alpha$ for $\alpha=-2.$ (b) Projection in the plane $x,y.$ This numerical elaboration shows that $P_4$ is the local attractor and $P_3^\pm$ are early-time attractors for this invariant set (actually they are saddles in the 6D phase space.)}
\end{center}
\end{figure}

\begin{figure}
\begin{center}
\includegraphics[height=2.5in,width=2.5in]{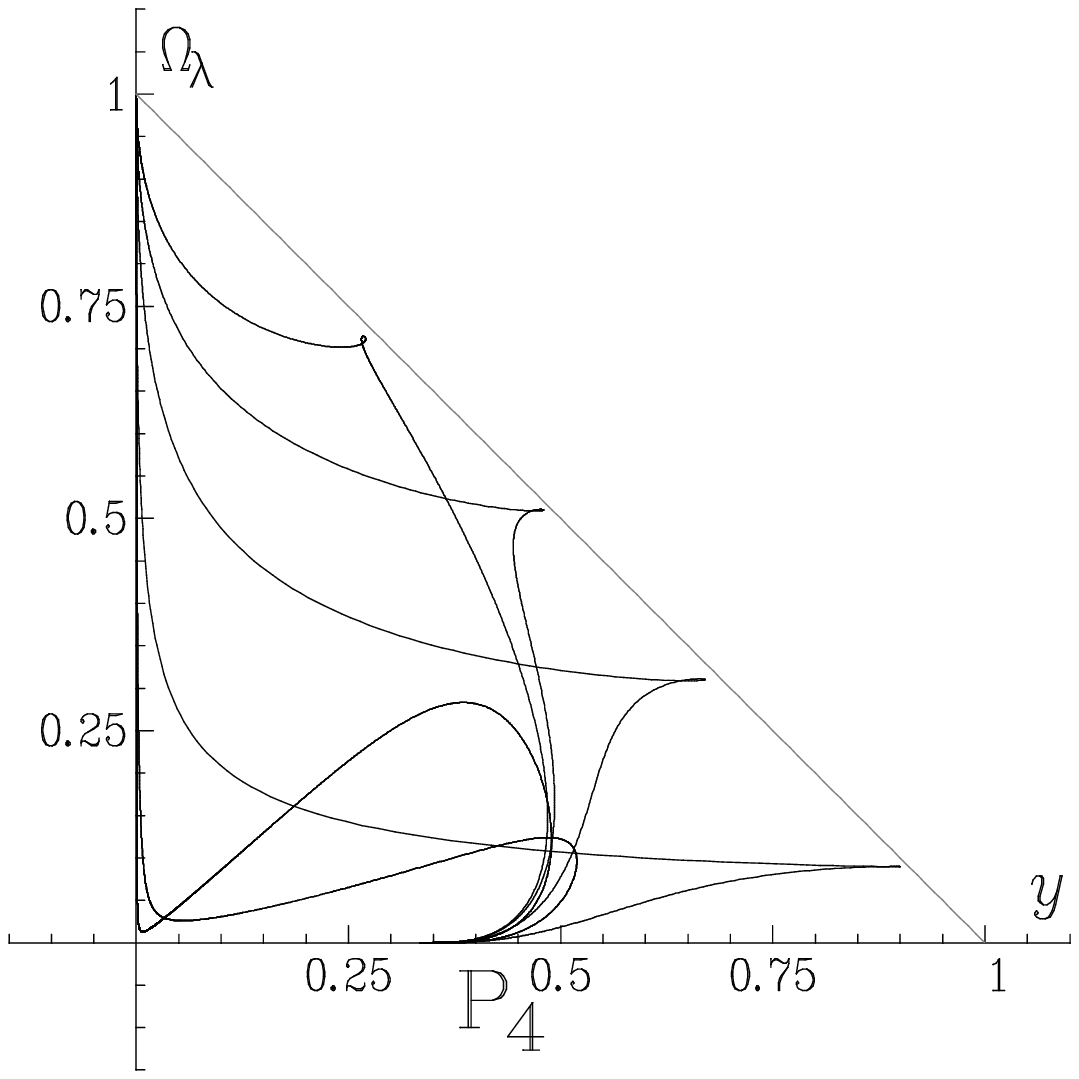}
\begin{center} (a) \end{center}
\includegraphics[height=2.5in,width=2.5in]{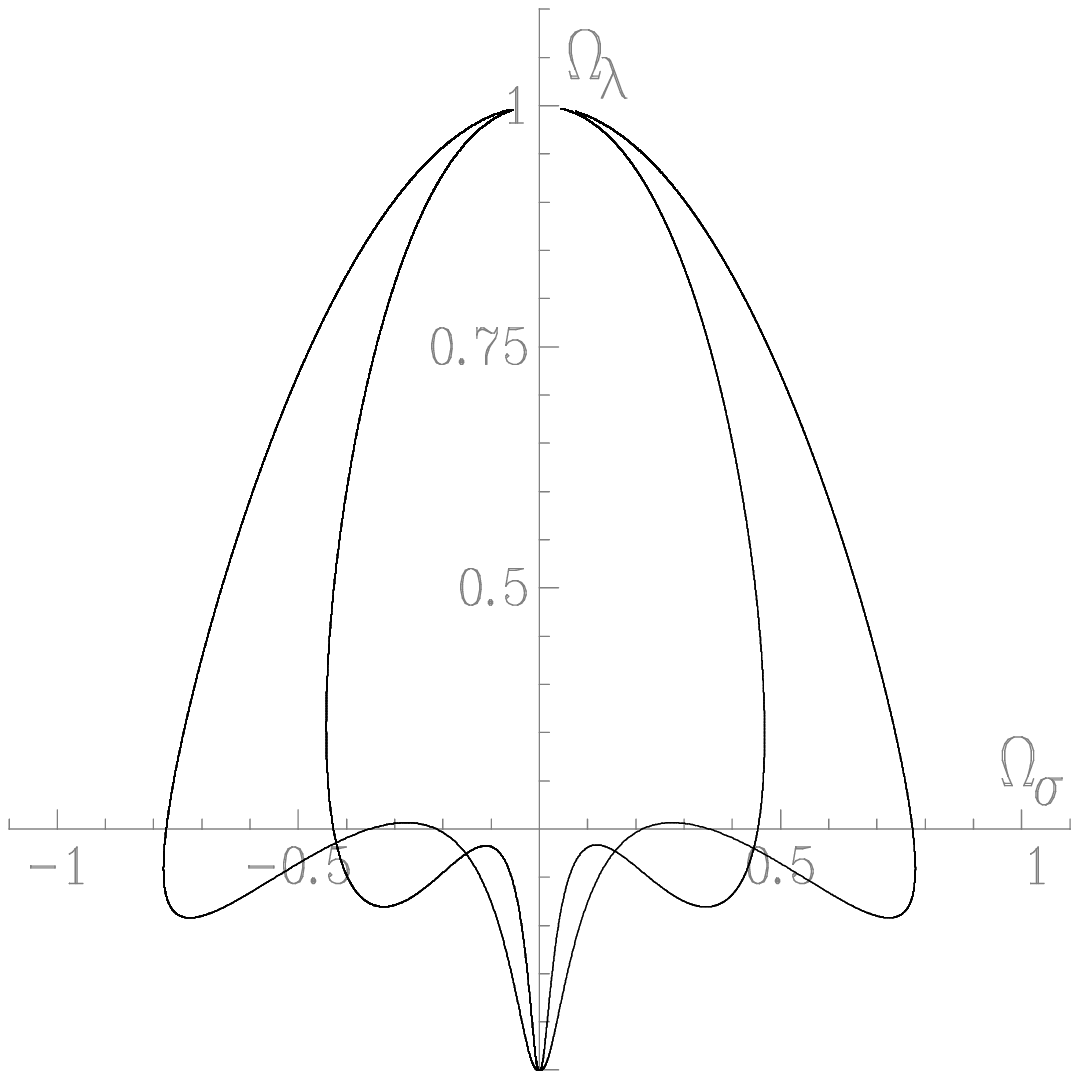}
\begin{center} (b) \end{center}
\caption{\label{P4attractorb} (a) Projection in the plane $y$-$\Omega_\lambda$ for the system
 \eqref{eqx}-\eqref{eqs} restricted to the invariant set $\Omega_m=\Omega_\sigma=0, s=2$   for the input function $f(s)=s+\alpha$ for $\alpha=-2.$ It is showed there the instability of the line $y+\Omega_\lambda=1$ and the stability of $P_4.$ In (b) are displayed some orbits projected in the plane $\Omega_\sigma, \Omega_\lambda.$ This figure illustrate that the universe evolves from a solution with non-standard 5D behavior to an isotropic solution ($P_4$).}
\end{center}
\end{figure}

In the figures \ref{P4attractora} and \ref{P4attractorb} are
presented some numerical integrations for the system
\eqref{eqx}-\eqref{eqs} for the input function $f(s)=s+\alpha$ for
$\alpha=-2.$ In this case the local attractor is the scalar
field-matter scaling solution ($P_4$). The point $P_5$ and $P_7$
coincides in which case they are non-hyperbolic (with saddle-type
behavior). We investigate the invariant set $s=2$ which contains
the relevant late-time attractor. Since the system is 6D the
dynamics is more richer than showed in the figures. However, as we
proved analytically, the universe evolves to an isotropic standard
scalar field-matter scaling solution.

\begin{figure}
\begin{center}
\includegraphics[height=2.5in,width=2.5in]{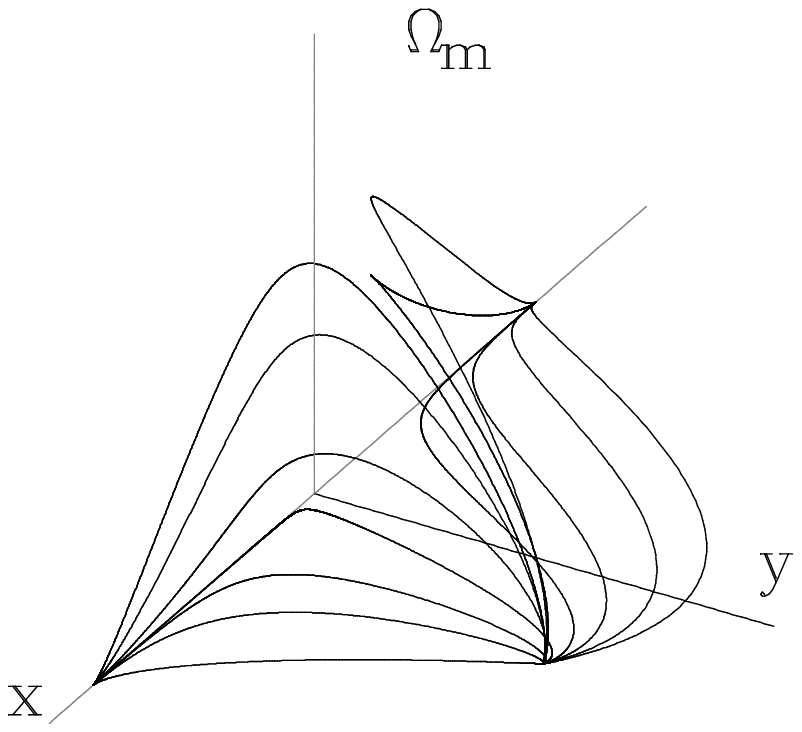}
\begin{center} (a) \end{center}
\includegraphics[height=2.5in,width=2.5in]{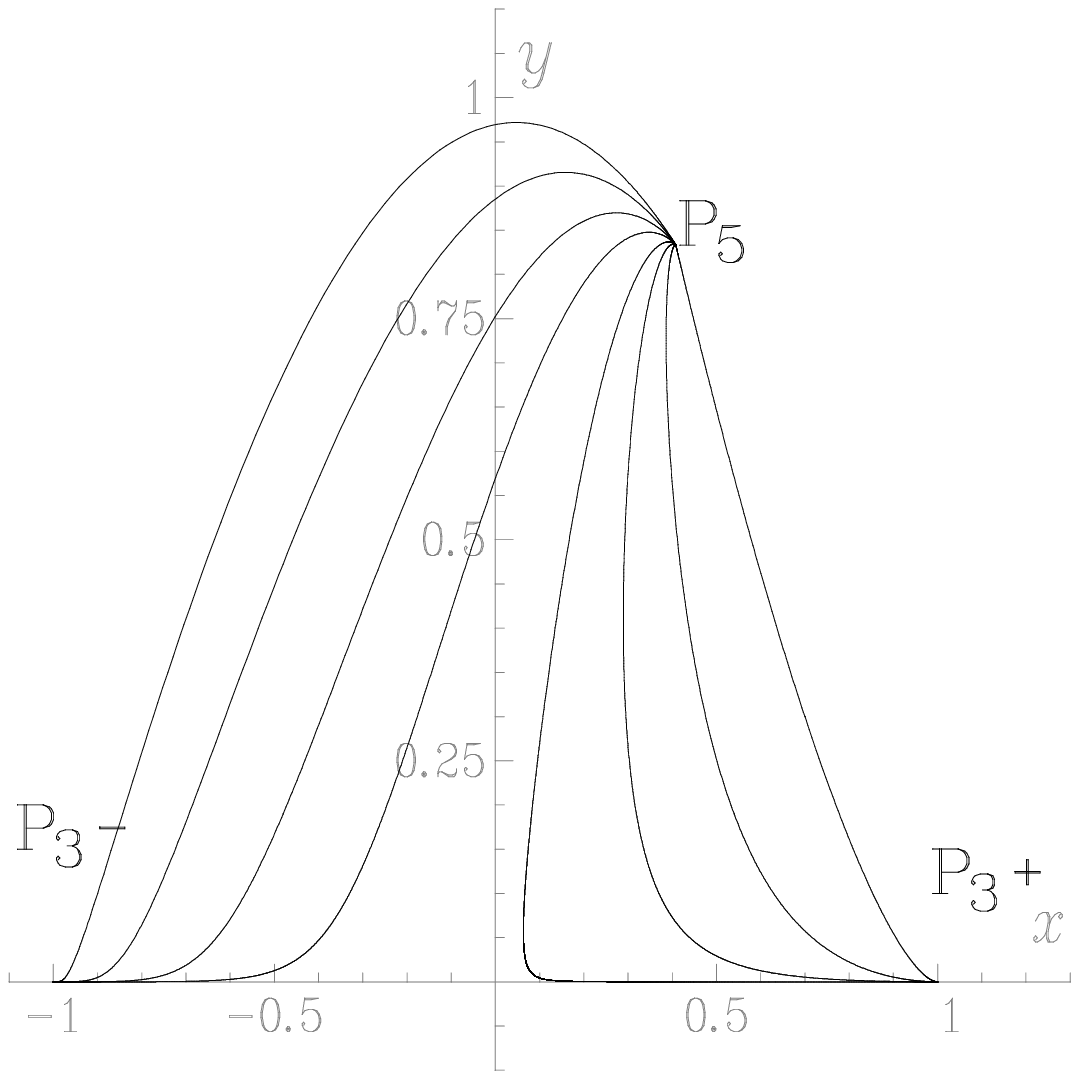}
\begin{center} (b) \end{center}
\caption{\label{P5attractora} (a) Some orbits in the invariant set $\Omega_\lambda=\Omega_\sigma=0, s=1$  of the system
 \eqref{eqx}-\eqref{eqs} for the input function $f(s)=s+\alpha$ for $\alpha=-1.$ (b) Projection in the plane $x,y,\Omega_m=0.$ This numerical elaboration shows that $P_5$ is the local attractor and $P_3^\pm$ are early-time attractors for this invariant set (actually they are saddles in the 6D phase space). $P_4$ and $P_7$ do not exist.}
\end{center}
\end{figure}

\begin{figure}
\begin{center}
\includegraphics[height=2.5in,width=2.5in]{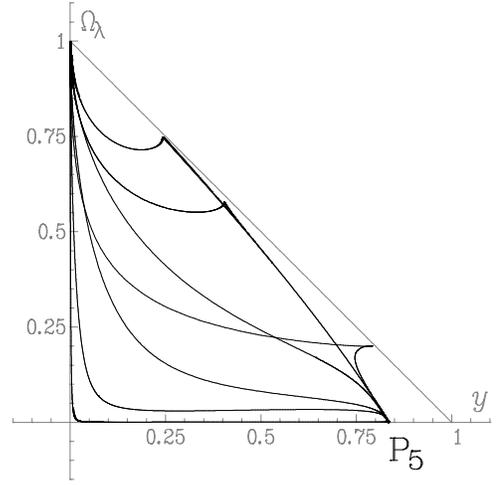}
\caption{\label{P5attractorb} Projection in the plane $y$-$\Omega_\lambda$ for the system
 \eqref{eqx}-\eqref{eqs} restricted to the invariant set $\Omega_m=\Omega_\sigma=0, s=1$   for the input function $f(s)=s+\alpha$ for $\alpha=-1.$ It is showed there the instability of the line $y+\Omega_\lambda=1$ and the stability of $P_5.$}
\end{center}
\end{figure}

In the figures \ref{P5attractora} and \ref{P5attractorb} are
presented some numerical integrations for the system
\eqref{eqx}-\eqref{eqs} for the input function $f(s)=s+\alpha$ for
$\alpha=-1.$ In this case the local attractor is the scalar
field-matter scaling solution ($P_5$). $P_4$ and $P_7$ do not
exist. We investigate the invariant set $s=1$ which contains the
relevant late-time attractor. As before, the system is 6D the
dynamics is more richer than showed in the figures, but, as we
proved analytically, the universe evolves to an isotropic standard
scalar field dominated solution.

\begin{figure}
\begin{center}
\includegraphics[height=2.3in,width=2.3in]{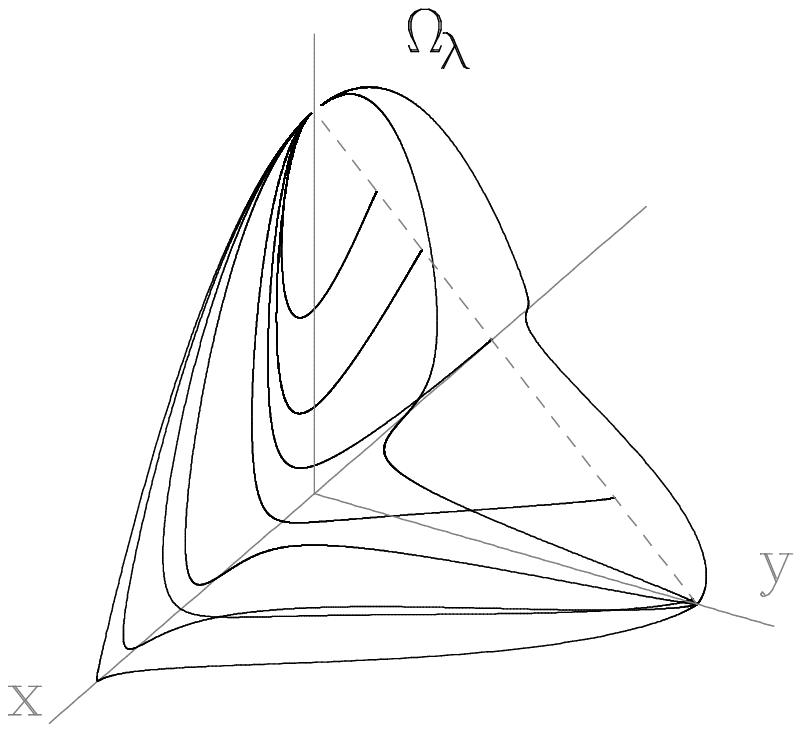}
\begin{center} (a) \end{center}
\includegraphics[height=2.3in,width=2.3in]{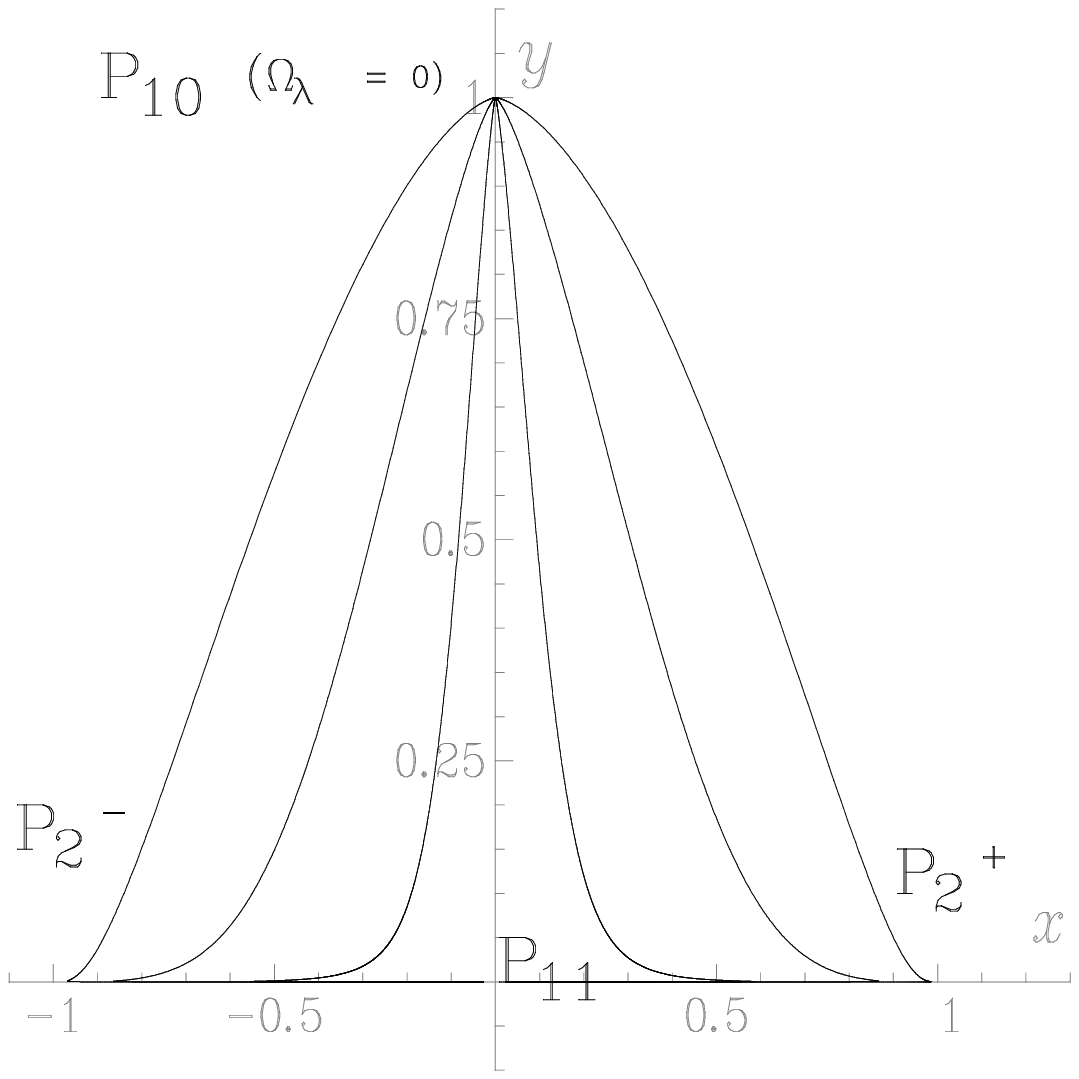}
\begin{center} (b) \end{center}
\includegraphics[height=2.3in,width=2.3in]{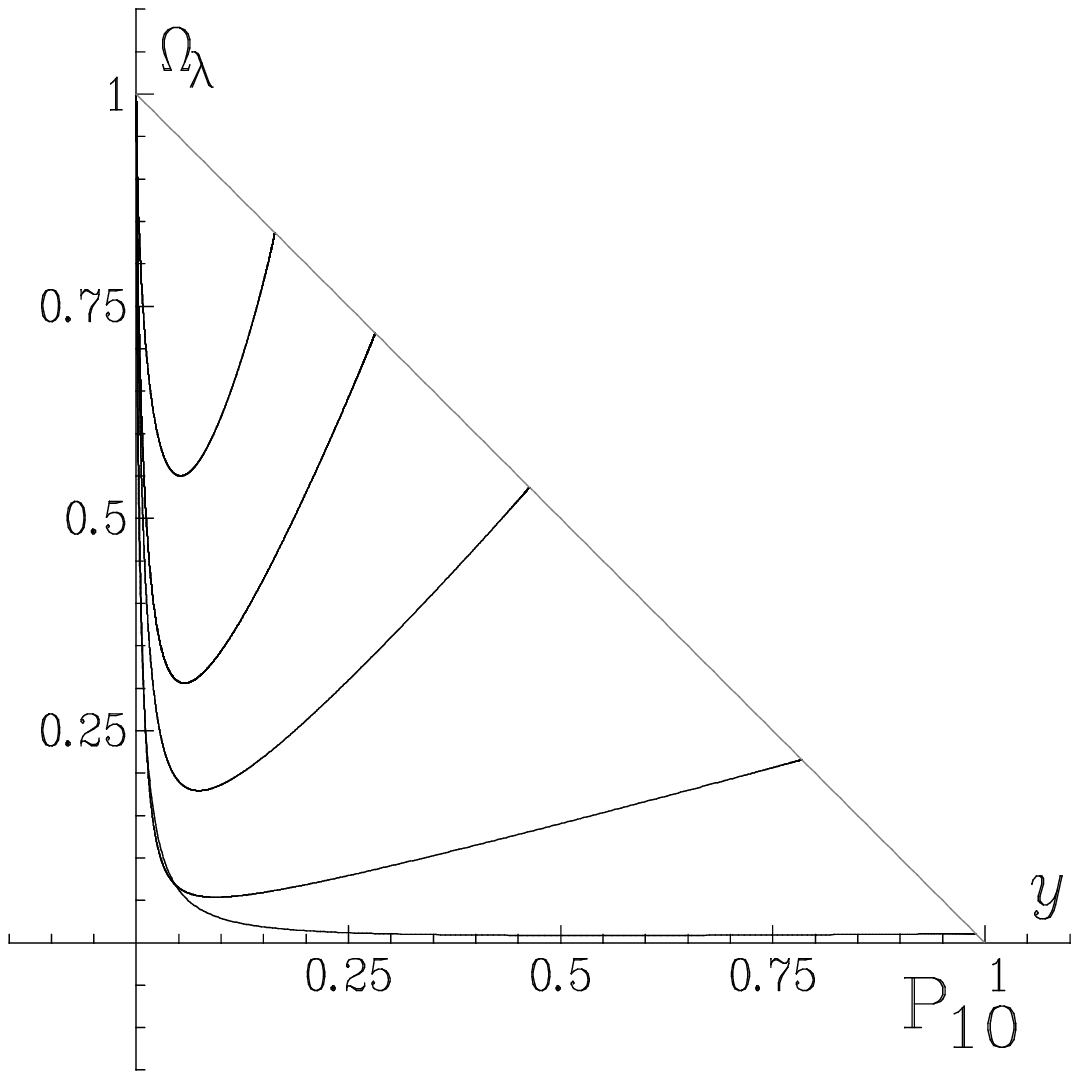}
\begin{center} (c) \end{center}
\caption{\label{P10attractora} (a) Some orbits in the invariant
set $\Omega_m=\Omega_\sigma=0, s=0$  of the system
 \eqref{eqx}-\eqref{eqs} for the input function $f(s)=s+\alpha$ for $\alpha=0.1.$
 (b) Projection in the plane $x,y.$ This numerical elaboration
 shows that $P_{10},\Omega_\lambda=0$ is the local attractor and $P_2^\pm$ are saddles.
 The local source in this invariant set is $P_{11}.$ (c) Projection of some orbits in the plane $y$-$\Omega_\lambda$
 for $\Omega_m=0$ suggesting that the line $y+\Omega_\lambda=1$ is a local attractor in the invariant
 set $\Omega_m=0.$ However they are saddles for the full dynamics.}
\end{center}
\end{figure}

In the figure \ref{P10attractora}(a) are displayed some orbits in the invariant
set $\Omega_m=\Omega_\sigma=0, s=0$  of the system
 \eqref{eqx}-\eqref{eqs} for the input function $f(s)=s+\alpha$ for $\alpha=0.1$
In \ref{P10attractora}(b) is presented the projection in the plane $x,y.$ This numerical elaboration
 shows that $P_{10},\Omega_\lambda=0$ is the local attractor and $P_2^\pm$ are saddles.
 The local source in this invariant set is $P_{11}.$ In \ref{P10attractora}(c) are drawn some orbits projected in the plane $y$-$\Omega_\lambda$
 for $\Omega_m=0$ suggesting that the line $y+\Omega_\lambda=1$ is a local attractor in the invariant
 set $\Omega_m=0.$ However they are saddles for the full dynamics.

\section{Results and discussion}

The  main results of this investigation can be summarized as follows. The singular point $P_{11}$ represents a Big-Bang singularity. It is non-hyperbolic with a 5D unstable manifold for $1<\gamma<2.$ In this case it is a local source as numerical simulations \ref{P4attractorb}, \ref{P5attractorb} and \ref{P10attractora}(c) suggest. For $0<\gamma<1$ it behaves as a saddle point.

In the general case $f(0)\in \mathbb{R},$ the solution $P_{10}$ with $\Omega_{\lambda}= 0$ is the late time de Sitter attractor without 5D-corrections for $f(0)>0$. To analyze their stability we have used the Center
Manifold Theory. Using this technique we have obtained an analogous center manifold for the origin (up to a variable-rescaling) to the one in  (41), (42) in \cite{Escobar2011}. This allow us to prove that $P_{10}$ with $\Omega_{\lambda}\in(0,1)$ are not late time attractors with 5D-modifications since they are always
saddle-like. This fact correlates with a transient primordial
inflation.  These results extent our results in \cite{Escobar2011} to the Bianchi models class.

The critical points $P^\pm_2$ are solutions dominated by the
kinetic energy of the scalar field, representing solutions
with an standard behavior ($\Omega_{\lambda}=0$), although non-hyperbolic, they behave as saddle-like
points in the phase space. Also are the critical point $P^\pm_3$ which are solutions dominated by the
kinetic energy of the scalar field. The critical points $P^\pm _6$ corresponding to scalar field-anisotropic scaling solutions
($\Omega_{\phi}\sim \Omega_{\sigma}$) are non-hyperbolic, however they behave as
saddle-like points since they have both nonempty stable and unstable manifolds. All these cosmological solutions represent transient states in the evolution of the universe.

The possible late-time stable solutions are
\begin{itemize}
\item The scalar field-matter scaling solution ($P_4$) provided $0<\gamma <\frac{4}{3},s^*<-\sqrt{3\gamma },f'\left(s^*\right)<0$ or $0<\gamma <\frac{4}{3},s^*>\sqrt{3\gamma },f'\left(s^*\right)>0.$
\item The scalar field-dominated solution ($P_5$) provided $0<\gamma \leq \frac{4}{3},-\sqrt{3 \gamma }<s^*<0,f'\left(s^*\right)<0$ or $\frac{4}{3}<\gamma \leq 2,-2<s^*<0,f'\left(s^*\right)<0$ or $0<\gamma \leq \frac{4}{3},0<s^*<\sqrt{3 \gamma },f'\left(s^*\right)>0$ or $\frac{4}{3}<\gamma \leq 2,0<s^*<2,f'\left(s^*\right)>0.$
\item The scalar field-dark radiation scaling solution ($P_7$) provided $\frac{4}{3}<\gamma \leq 2,s^*<-2,  f'\left(s^*\right)<0$ or $\frac{4}{3}<\gamma \leq 2,s^*>2,  f'\left(s^*\right)>0.$
\item The de Sitter solution $P_{10}$ with $\Omega_\lambda=0$ provided $f(0)$ is a real (finite) positive number, i.e., $f(0)>0.$
\end{itemize}
The main difference with respect to our analysis in \cite{Escobar2011} is that there are possible scalar field-dark radiation scaling late time solutions ($P_7$) for a wide region in the parameter space (see figure \ref{parameter} for a numerical elaboration).

Our results are quite general and apply to the scalar fields potentials presented in table I in the reference \cite{Escobar2011}. In the particular case of a scalar field with potential
$V=V_{0}e^{-\chi\phi}+\Lambda$ it can be proved, in a similar way as in \cite{Escobar2011}, that for $\chi<0$ the de Sitter solution is asymptotically stable. However, for $\chi>0$ the de Sitter solution is unstable (of saddle type).
We omit the calculation here because it is straightforward and the result is consistent with our previous results in \cite{Escobar2011}. Instead of considering classical potentials discussed in the literature we have considered a toy model for illustrating our analytical findings. In this case we have investigated a scalar field
with potential $V(\phi)=V_0 \left(-\frac{1}{W\left(-e^{-\phi  \alpha
^2-1}\right)}\right)^{\frac{1}{\alpha }}$ where
$W(z)$ is the special function `ProductLog' that gives the
principal solution for $w$ in $z=we^w.$ This potential corresponds to the $f$-deviser $f(s)=s+\alpha.$

In this paper we are mainly interested in the late-time dynamics. For the above particular example the late-time dynamics can be summarized as follows.

\begin{itemize}
\item The scalar field-matter scaling solution ($P_4$) is a  late-time attractor provided  $0<\gamma <\frac{4}{3},\alpha<-\sqrt{3\gamma }.$
\item The scalar field-dominated solution ($P_5$) is a late-time attractor provided  $0<\gamma \leq \frac{4}{3},-\sqrt{3 \gamma}<\alpha<0$ or $\frac{4}{3}<\gamma \leq 2,-2<\alpha<0.$
\item The scalar field-dark radiation scaling solution ($P_7$) is a late-time attractor provided  $\frac{4}{3}<\gamma \leq 2,\alpha<-2.$
\item The de Sitter solution $P_{10}$ with $\Omega_\lambda=0$ is the late-time attractor provided $\alpha>0.$
\end{itemize}

\section{Conclusions}

In the present paper we have investigated the phase space of the
Randall-Sundrum braneworlds models with a self-interacting scalar
field trapped in a Bianchi I brane with arbitrary potential.

From our numerical experiments we claim that $P_{11}$ is associated with the
Big Bang singularity type. The numerical investigations performed in this paper suggest
that it is in general the past attractor in the phase space of the
Randall-Sundrum cosmological models.

Using the center manifold theory we have obtained sufficient conditions for the
asymptotic stability of de Sitter solution with standard 4D behavior.
We have proved, using the center manifold theory, that there are not late time de Sitter attractors
with 5D-modifications since they are always saddle-like. This fact
correlates with a transient primordial inflation. We have obtained sufficient conditions on the potential for the stability of the  scalar field-matter scaling solution, the scalar field-dominated solution, and the scalar field-dark radiation scaling solution. We illustrate our analytical findings using a simple $f$-deviser as a toy model.
All these results are generalizations of our previous results obtained for FRW branes in \cite{Escobar2011}.

\section*{Acknowledgments}
This work was partially supported by PROMEP, DAIP and by CONACyT, Mexico, under grant
167335; by MECESUP FSM0806, from Ministerio de Educaci\'on de Chile; and by the National
Basic Science Program (PNCB) and Territorial CITMA Project (no. 1115), Cuba. DE, CRF
and GL wish to thank the MES of Cuba for partial financial support of this investigation. YL
is grateful to the Departamento de F\'isica and the CA de Gravitaci\'on y F\'sica Matem\'atica for
their kind hospitality and their joint support for a postdoctoral fellowship.


\end{document}